\documentclass[12pt]{article}

\def\TL{\hfil$\displaystyle{##}$}
\def\TR{$\displaystyle{{}##}$\hfil}

\def\TT{\hbox{##}}
\def\seqalign#1#2{\vcenter{\openup1\jot
  \halign{\strut #1\cr #2 \cr}}}

\def\mop#1{\mathop{\rm #1}\nolimits}

\def\sech{\mop{sech}}

\def\Vol{\mop{Vol}}

\def\overleftrightarrow#1{\vbox{\ialign{##\crcr
     $\leftrightarrow$\crcr\noalign{\kern-0pt\nointerlineskip}
     $\hfil\displaystyle{#1}\hfil$\crcr}}}

\def\lsim{\mathrel{\mathstrut\smash{\ooalign{\raise2.5pt\hbox{$<$}\cr\lower2.5pt\hbox{$\sim$}}}}}
\def\gsim{\mathrel{\mathstrut\smash{\ooalign{\raise2.5pt\hbox{$>$}\cr\lower2.5pt\hbox{$\sim$}}}}}

\def\sqr#1#2{{\vcenter{\vbox{\hrule height.#2pt
         \hbox{\vrule width.#2pt height#1pt \kern#1pt
            \vrule width.#2pt}
         \hrule height.#2pt}}}}

\def\href#1#2{#2}

\def\lbldef#1#2{\expandafter\gdef\csname #1\endcsname {#2}}
\def\eqn#1#2{\lbldef{#1}{(\ref{#1})}%
\begin{equation} \eqalign{#2} \label{#1} \end{equation}}
\def\eqalign#1{\vcenter{\openup1\jot
    \halign{\strut\span\TL & \span\TR\cr #1 \cr
   }}}

\begin{document}
\pagestyle{plain}
\setcounter{page}{1}
\begin{titlepage}

\begin{flushright}
PUPT-2105 \\
hep-th/0312321
\end{flushright}
\vfil

\begin{center}
$\,$\lower5pt\hbox{\huge String creation and cosmology}%
\large\footnote{Contribution to the proceedings of QTS3, held at the University of Cincinnatti from September 10 to 14, 2003.}%
\end{center}

\vfil
\begin{center}
{\large Steven S. Gubser}
\end{center}

$$\seqalign{\span\TL & \span\TT}{
& Joseph Henry Laboratories, Princeton University, Princeton, NJ 08544  }$$
\vfil

\begin{center}
{\large Abstract}
\end{center}

\noindent
I argue that string creation may have played a role in reheating the universe after inflation.  For strings in four dimensions that arise from branes wrapping cycles in the extra dimensions, estimates from effective field theory show that the string tension need only fall a couple of orders of magnitude below the Planck scale in order for string creation to extract a significant fraction of the energy in coherent motion of the inflaton field.  I also comment on a special four-dimensional background which involves only Neveu-Schwarz fields and offers the possibility of studying closed string creation on the worldsheet.

\vfil
\begin{flushleft}
December, 2003
\end{flushleft}
\end{titlepage}
\newpage

\section{Introduction}
\label{INTRODUCTION}

As the experimental data on the large-scale structure and early history of the universe improve, there has been increased interest in seeing how ideas from string theory may fit in to understanding cosmology at the earliest of times.  On the whole, efforts in this direction have focused on either finding a home for inflation in string theory (for instance, \cite{Alexander,kklmmt}) or replacing it with something quite different but string-inspired (for instance, \cite{bv,Veneziano,kosst}).  Important as it is to establish in a fundamental theory of gravity either a sound underpinning for inflation or a solid alternative to it, my focus here will be instead on the possibility of string physics playing an important role just after inflation has ended.  The characteristic scale of string theory is usually closely to the Planck mass, considerably higher than a typical inflaton mass scale.  However, it turns out that objects known as ``tensionless strings'' \cite{WittenComments,AndyOpen,ghSmall,swComments} can be produced copiously from coherent motion of the inflaton field.  Without significant fine-tuning, it is possible that creation of such strings can extract a significant fraction of energy from the inflaton, thus providing an alternative to preheating \cite{bt,kls,zhs}.

String creation is interesting in its own right, but on the whole it is ill-understood.  The best studied examples of it are decaying D-brane solutions \cite{senRoll,AndyCreate,stromingerEtAl,maldacenaEtAl}, where boundary CFT methods are available.  We will rely instead on estimates from effective field theory, employing a method of steepest descent to extract Bogliubov coefficients from approximate wave equations for massive string modes.  This method was developed in \cite{ml,gCreate}, and its application to preheating was considered in \cite{gCreate}.  It would be highly desirable to improve our ability to compute string creation amplitudes: this is an important feature of time-dependent backgrounds in string theory, and it is relevant to almost all the stringy proposals for the physics of the early universe.

The organization of this paper is as follows.  In section~\ref{CREATE} we will briefly summarize the results of the steepest descent analysis of string creation.  In section~\ref{PREHEAT} we will explain in some detail how this analysis may be applied to provide an alternative to preheating.  In section~\ref{DILATON} we will remark on a special background in which a worldsheet treatment of closed string creation may be possible.

The results summarized here are based mainly on \cite{gCreate}, except that section~\ref{DILATON} is based on ongoing work with J.~Friess and I.~Mitra.

\section{Estimating string creation}
\label{CREATE}

The density of states in tree-level string theory is exponential:
 \eqn{DOS}{
  {dN \over dE} \propto E^\gamma e^{E/T_H} \qquad
   T_H = {1 \over 2\pi \sqrt{\alpha' c_\perp/6}} \,,
 }
where $T_H$ is the Hagedorn temperature and $\gamma$ depends on details of the string theory in question.  The quantity $c_\perp$ is the central charge for transverse modes: $c_\perp = 12$ for the type II superstring.  The exponential density of states \DOS\ is in contrast with the power-law behavior one encounters for point-particle theories.  The states that dominate \DOS\ are highly excited strings.  String interactions are not taken into account in \DOS.  $T_H$ appears to be a temperature at which some phase transition occurs.  Above the phase transition strings themselves may not be a good description of the physics.  There is an extensive literature on this so-called Hagedorn transition (see for example \cite{Kogan,aw,ggkrw}), and string interactions do enter into the discussion in important ways.  For the purposes of the present discussion, we will not need to consider interactions explicitly.

The on-shell constraints for quantum states of a string include $L_0 |\psi_{\rm phys}\rangle = 0$, which leads to an equation roughly of the form
 \eqn{Onshell}{
  \ddot\chi + \omega(t)^2 \chi = 0 \,,
 }
where $\omega(t)^2$ depends on the excitation level and the momentum: in a flat background, $\omega^2 = k^2 + N/\alpha'$, where $N$ is the level (up to factors of $2$ that depend somewhat on details).  In a cosmological background, a reasonable first guess is 
 \eqn{OmegaForm}{
  \omega(t)^2 = k^2 + a(t)^2 N/\alpha' \,,
 }
where $a(t)$ is the scale factor and $t$ is conformal time.  One would of course like to improve upon \Onshell: this is one place where input from a worldsheet treatment is needed.  The form \OmegaForm\ is predicated on working with the string frame metric, where the parameter $\alpha'$ is a constant length squared.

The steepest descent method, developed in this context in \cite{ml,gCreate}, allows us to crudely estimate the Bogliubov coefficients for \Onshell: if $|\beta| \ll 1$, then
 \eqn{BetaEstimate}{
  \beta \approx {i\pi \over 3}
   \exp\left( -2i \int_{-\infty}^4 dt \, \omega(t) \right)
   \exp\left( -2i \int_r^{t_*} dt \, \omega(t) \right) \,,
 }
where $t_* = r - i\mu$ is the location of the zero of $\omega(t)^2$ in the lower half plane which is closest to the real axis, and $r$ and $\mu$ are real.  The last of the three factors in \BetaEstimate\ is the most important: it leads to exponential suppression of $|\beta|^2$ when $\omega$ is large.  Assuming that $\omega(t)$ is well-approximated by an elliptical arc over the contour from $r$ to $t_*$, one has
 \eqn{MuApprox}{
  |\beta|^2 \approx \left( {\pi \over 3} \right)^2 
   e^{-\pi \mu \omega(r)} \,.
 }
From this one can estimate the total number of strings created:
 \eqn{TotalNumber}{
  N_{\rm tot} &= \int^\infty d\omega \, {dN \over d\omega} |\beta|^2
    \sim \int^\infty d\omega \, e^{\omega/T_H} e^{-\pi\mu\omega} \,.
 }
Evidently, this total number {\sl converges} precisely if
 \eqn{MuBound}{
  \pi\mu T_H > 1 \,.
 }
If this bound is violated, it's somewhat like passing to a temperature above the Hagedorn temperature: clearly, interactions and/or back-reaction must then be involved in rendering the total energy passing into strings finite.

The punch-line of the next section will be that, for non-perturbative strings, it is possible without appreciable fine-tuning of parameters to violate \MuBound\ just after inflation.  If this happens, then strings must indeed play an important role in reheating.  Moreover, it would be highly excited strings that would make the dominant contribution, so the effects of strings would be quite unlike anything in a point-particle theory---hence characteristic of string theory.

Needless to say, it would be very desirable to learn what happens in time-dependent backgrounds when the number or energy of strings created diverges.  Decaying D-branes may be the simplest laboratory in which to attack this question: see for example \cite{gir}.

\section{Extracting energy from inflaton oscillations}
\label{PREHEAT}

Let us start with a brief overview of the theory of preheating \cite{bt,kls,zhs}, which will be a useful point of comparison when considering post-inflationary string creation.
 \begin{itemize}
  \item At the end of inflation, the universe is cold and empty.
  \item The inflaton $\phi$ is oscillating around its minimum with an
amplitude assumed to be comparable to the Planck scale.
  \item These coherent oscillations may die off via perturbative
production of particles.  This is ordinary reheating. 
  \item Preheating can occur if another {\it boson} has a coupling $g \phi^2\chi^2$ to the inflaton.
  \item The effective mass of $\chi$ varies: $m_{\rm eff}^2 = m_0^2 + g\phi^2$.  This means parametric resonance can occur and give
exponentially growing occupation numbers for $\chi$.
  \item Preheating is ``efficient'' iff $m_{\rm eff}^2$ varies by a large factor from its minimum to its maximum: $m_0^2$ needs to be a small contribution to the mass.
 \end{itemize}
The last point appears contrived, especially when it is stated more precisely, as we will do shortly.  ``Efficient'' in this context means that a significant fraction of the energy in the coherent inflaton oscillations should be extracted via the parametric resonance.  We may then ask, if string states could be produced in a similar way, could they play a role in reheating?

We have already noted in the introduction that the string scale is usually only slightly smaller than the four-dimensional Planck scale, $M_{\rm Pl,4} = (8\pi G_N)^{-1/2} \sim 10^{18} \, {\rm GeV}$, while a typical value for the inflaton mass (in power-law inflation) is $m_{\rm inflaton} \sim 10^{13} \, {\rm GeV}$.  It will not come as a surprise that this prevents copious production of fundamental strings, unless we make somewhat unnatural assumptions.  Although our eventual aim is to consider production of non-perturbatively constructed strings which are tensionless at some point in moduli space, it will be instructive first to consider the production of fundamental strings in slightly more detail; also, this discussion will tie in with the special background discussed in section~\ref{DILATON}.

The first goal in assessing a model of preheating is to find the regime of parameters where preheating is efficient.  The analogous goal here is to find where the bound \MuBound\ is first violated.  When this happens, the particle occupation numbers of the individual string modes should still be small: if they are not, then according to \TotalNumber, the number of strings created will diverge exponentially.  Thus we have no need of parametric resonance and can continue to use the steepest descent method.

The onshell condition in string metric is, approximately, $\omega^2 = k^2 + m^2 |g_{tt}^{str}|$, where $t$ is conformal time.  The string metric is related to the Einstein metric by
 \eqn{EinString}{
  ds_{4E}^2 = e^{\gamma\varphi/M_{\rm Pl,4}} ds_{str}^2 = 
   a(t)^2 (-dt^2 + d\vec{x}^2) \,,
 }
where $\varphi$ is the dilaton and $\gamma$ is a constant of order unity.  Thus
 \eqn{ModifiedOmega}{
  \omega(t)^2 = k^2 + e^{\gamma \varphi(t)/M_{\rm Pl,4}} a(t)^2 
   {N \over \alpha'} \,.
 }
The dilaton is probably not the same field as the inflaton; however, it seems reasonable to assume some finite overlap, so when the inflaton is oscillating at the end of inflation, the dilaton is too.  Thus $\omega(t)^2$ oscillates between slowly varying extremes:
 \eqn{OmegaApproximate}{
  \omega(t)^2 &= 
   {\omega_{\rm max}^2 + \omega_{\rm min}^2 \over 2} + 
   {\omega_{\rm max}^2 - \omega_{\rm min}^2 \over 2} 
    \cos \Omega t  \cr
  \omega_{\rm min}^2 &= 
   k^2 a(t)^2 + e^{\gamma\varphi_{\rm min}/M_{\rm Pl,4}}
   {N \over \alpha'}  \cr
  \omega_{\rm max}^2 &= k^2 a(t)^2 + e^{\gamma\varphi_{\rm max}/M_{\rm Pl,4}}
   {N \over \alpha'} \,.
 }
The zeroes of $\omega(t)^2$ are
 \eqn{tPoles}{
  t^{\pm}_{*,n} = {2\pi n \over \Omega} \pm {i \over \Omega}
   \log {\omega_{\rm max} + \omega_{\rm min} \over 
     \omega_{\rm max} - \omega_{\rm min}} \,,
 }
and it is plausible to assume that $t_* = t^-_{*,0} = r - i\mu$ makes
the biggest contribution to string production.  String production is finite if
 \eqn{MuBoundAgain}{
  \pi \mu T_H > 1 \,,
 }
which translates approximately to
 \eqn{TensionIneq}{
  \sqrt{\tau_{\rm min}} \gsim m_{\rm inflaton} \,,
 }
where $\tau_{\rm min}$ is the minimum string tension as measured in Einstein frame.  

If the bound \MuBoundAgain\ is violated, then what must happen
is that a significant fraction of the energy of the inflaton
oscillations passes into excited string states, and the back-reaction
slows down the inflaton.  This would correspond to the idea of efficient preheating.  Note however that $\omega_{\rm min} \sim \sqrt{\tau_{\rm min}}$ and $\Omega \sim m_{\rm inflaton}$, so for \TensionIneq\ to occur, the string tension as measured in Einstein frame must dip temporarily below $m_{\rm inflaton}$.  This is possible, but it seems like asking a lot in conventional string compactifications.  The upshot, then, is that dimensional analysis didn't fail, and we have proposed a contrived scenario.

Let's compare with conventional preheating, where it is useful to define ``Mathieu parameters'' $A$ and $q$ as follows:
 \eqn{MathieuParameters}{
  A + 2q = \left( {\omega_{\rm max} \over 2\Omega} \right)^2 \,, \quad
  A - 2q = \left( {\omega_{\rm min} \over 2\Omega} \right)^2 \,.
 }
A ``broad resonance'' occurs in the standard theory when $2q <
A < 2q + \sqrt{bq}$ for some order one constant $b$: this leads to efficient particle production.  On the other hand, efficient string production occurs when $2q < A < 2xq$ for some $x>1$.  So string production occurs in a qualitatively bigger wedge of parameter space, albeit still a small one.

Another problem with proposing that fundamental strings might have been created after inflation is that they can wind around the extra dimensions.  In phenomenologically interesting string compactifications, there usually is a non-trivial first homotopy group on the compactification manifold, so such windings can be topologically stable.  Moreover, the density of states of wound strings is smaller only by roughly a constant factor compared to unwound strings, and this constant factor is at most a couple of orders of magnitude in conventional setups.  Wound strings would act as dark matter, or perhaps they could even have visible sector effects like fractional charge.  Just the dark matter consideration is discouraging: a naive estimate suggests that wound strings would overclose the universe today by something close to a factor of $10^{17}$.

Considering non-perturbatively constructed strings in place of fundamental strings alleviates all the difficulties so far discussed.  These strings would arise as branes wrapped on a shrinking cycle in the extra dimensions.  Consider for example D3-branes on a shrinking $S^2$.  The $S^2$ may become small only at a point on the compactification manifold.  This avoids the difficulty of having wound strings, because there is a strong energy barrier against wrapped branes wandering away from the location of the shrinking cycle.  The parameter controlling the size of the $S^2$ is in fact complex: it is
 \eqn{ComplexVolume}{
  \varphi = \int_{S^2} (J_2 + i B_2) \,,
 }
where $J_2$ is the Kahler form on the extra dimensions (assumed to form a complex manifold) and $B_2$ is the Neveu-Schwarz two-form potential of string theory.  Starting from the DBI action for D-branes, one arrives at
 \eqn{DBIAction}{
  S = -\tau_{D3} \int d^4 \xi \sqrt{G_{\mu\nu}+B_{\mu\nu}} + \ldots
 }
for a D3-brane wrapped on the $S^2$.  Using $\tau_{D3} \sim 1/(\alpha'^2 g_s)$, one obtains
 \eqn{tauEff}{
  \tau_{\rm eff} \approx M  |\varphi| 
   \qquad\hbox{with}\qquad M \sim {M_{\rm Pl,4} \over g_s} \,,
 }
where $\varphi$ has been rescaled to be a canonically normalized complex scalar field.  We have derived \tauEff\ using a particular construction of light strings in four dimensions, but it is in fact typical for a variety of non-perturbative constructions, many of which are related to one another by string dualities.

Let us now consider the situation where after inflation, various moduli of string theory are rolling around, presumably before stabilizing in a preferred vacuum.  The scalar $\varphi$ in particular rolls around the complex plane, and if it comes close to the origin, then the light strings we have discussed become nearly tensionless.  Based on our earlier discussion, we may expect appreciable string creation at such a moment.  To estimate it, suppose $\varphi(t) \approx \varphi_0 + \dot\varphi t$, where $\varphi_0$ is the closest approach to the origin on the trajectory $\varphi(t)$.  Then
 \eqn{tauVaries}{
  \tau_{\rm eff} \approx M |\varphi_0 + \dot\varphi t|
   = M \sqrt{|\varphi_0|^2 + (|\dot\varphi| t)^2} \,.
 }
To estimate the amount of string creation, we want to find a complex solution $t_*$ to $\omega(t)^2 = 0$.  For highly excited strings, it is equivalent to solve $\tau_{\rm eff}=0$, which gives
 \eqn{SingSolve}{
  t_* \equiv r-i\mu = -i |\varphi_0|/|\dot\varphi| \,.
 }
Thus the convergence criterion on the total number of strings produced is
 \eqn{ConvergenceC}{
  \pi \mu T_H \sim 
   {|\varphi_0| \over |\dot\varphi|}\sqrt{M|\varphi_0|} \gsim 1 \,.
 }
The minimum tension for the light strings is $\tau_{\rm min} \approx M|\varphi_0|$.  We can estimate that $\dot\varphi \sim M_{\rm Pl,4} m_{\rm inflaton}$: the $\varphi$ oscillations are of amplitude $M_{\rm Pl,4}$ and of frequency $m_{\rm inflaton}$.  The condition \ConvergenceC\ becomes
 \eqn{BoundSummary}{
  1 &\lsim {\sqrt{M \varphi_0^3} \over \dot\varphi}
   = {\tau_{\rm min}^{3/2} \over M \dot\varphi}
   \sim {\tau_{\rm min}^{3/2} \over M M_{\rm Pl,4} m_{\rm inflaton}}
   \sim 10^4 
    \left( {\tau_{\rm min} \over M_{\rm Pl,4}^2} \right)^{3/2} \,,
 }
or equivalently,
 \eqn{FinalBound}{
  \sqrt{\tau_{\rm min}} \gsim {1 \over 20} M_{\rm Pl,4} \,,
 }
where we have assumed $M \sim 10 M_{\rm Pl,4}$ and $m_{\rm inflaton} \sim 10^{-5} M_{\rm Pl,4}$.

The estimates leading to \FinalBound\ are inevitably fairly crude as long as we have not specified what the inflaton is and how much it overlaps with the scalar $\varphi$ controlling the effective string tension.  The result \FinalBound\ is rather different from what we had for fundamental strings, and the reason is that we almost
have a discontinuity in the first derivative of $\omega(t)^2 \propto \tau_{\rm eff} \sim M|\varphi| \approx M |\dot\varphi t|$.
 
Evidently, the non-perturbatively constructed strings do not
have to dip very far below the Planck scale in order to violate this
convergence criterion!  So their participation in reheating is quite plausible.  One would still have to answer the question of how dense in moduli space the loci of tensionless strings are.  Quantifying this is technically difficult, but the fact that the loci are typically real codimension two is encouraging.

\section{Backgrounds involving only Neveu-Schwarz fields}
\label{DILATON}

Let us now consider a background which might be more amenable to a worldsheet treatment.\footnote{This section represents work in progress with J.~Friess and I.~Mitra.}  As noted earlier, it is desirable to explore the possibilities for such vacua, since one would eventually want to go beyond the effective field theory method of estimating string creation that we described in section~\ref{CREATE}.  Essentially all string theory compactifications give rise to the following terms in the effective action (written in four-dimensional string frame):
 \eqn{StringFrameAction}{
  S &= {1 \over 2\kappa_4^2} \int d^4 x \, \sqrt{G} e^{-\phi}
   \left[ R + (\partial\phi)^2 - {1 \over 2} H_3^2 \right]  \cr
  e^{-\phi} &= {\Vol M_6 \over (2\pi\sqrt{\alpha'})^6}
   {1 \over g_s^2} \qquad
  2\kappa_4^2 = 2\pi\alpha' \,.
 }
In four-dimensional Einstein frame, defined by $ds_E^2 = e^{-\phi} ds_{str}^2$, we have instead
 \eqn{EinsteinFrameAction}{
  S = {1 \over 2\pi\alpha'} \int d^4 x \, \sqrt{g}
   \left[ R - {1 \over 2} (\partial\phi)^2 - {1 \over 2} 
    e^{2\phi} (\partial\chi)^2 \right] \,,
 }
and the fundamental string tension as measured in Einstein frame is $\tau_{\rm eff} = e^\phi/2\pi\alpha'$.  Except in type I string theory, $H_3$ is the four-dimensional part of the field strength of the Neveu-Schwarz two-form, and $\chi$ is defined via the equation $d\chi = e^{-\phi} *H_3$.

The scalar $\phi$ and the axion $\chi$ parametrize the upper half plane.  As a first guess, we might suppose that there are FRW solutions where $\chi + i e^{-\phi}$ approximately describes a semi-circular geodesic in the upper half plane (with the usual metric), and that by making $g_s$ small, we can avoid significant back-reaction of the scalars on the geometry.  Solving the scalar equations of motion in a background assumed to be flat Minkowski space, one finds
 \eqn{FlatChiPhi}{
  \chi + i e^{-\phi} = e^{-\phi_0} \left(
   \tanh {t \over t_0} + i \sech {t \over t_0} \right) \qquad
  \tau_{\rm eff} = {e^{\phi_0} \cosh t/t_0 \over 2\pi\alpha'} \,,
 }
where $\phi_0$ and $t_0$ are integration constants.  The usual convergence criterion for the total number of strings created is $t_0 \sqrt{\tau_{\rm eff}(0)} \gsim 1$, or, more explicitly,
 \eqn{StringsCreated}{
  e^{\phi_0} \gsim {2\pi\alpha' \over t_0^2} \,.
 }
It would appear that by making $\phi_0$ very negative (corresponding to extremely weak string coupling), with $t_0$ large compared to $\sqrt{\alpha'}$ but fixed, we can get the bound in \StringsCreated\ to fail.  This is a first indication that this type of solution is an interesting laboratory for string creation.  It is nice that the wave equation for string modes is approximately a Mathieu equation, and that in the early and late time limits it is possible to construct infinite order adiabatic in and out vacua.

In fact, it is not difficult to promote the approximate solution described above to an exact solution of the equations of motion following from the two-derivative action:
 \eqn{FullSolution}{
  H_3 &= e^\phi * d\chi = h dx^1 \wedge dx^2 \wedge dx^3 \qquad
   e^\phi = e^{\phi_0} \sec ht  \cr
  ds_{4E}^2 &= -e^{\phi_0} (\sec^2 ht)
   \left( {\cos {ht \over 2} + \sin {ht \over 2} \over
    \cos {ht \over 2} - \sin {ht \over 2}} \right)^{\sqrt{3}} dt^2 + 
   e^{-\phi_0 \over 3}
   \left( {\cos {ht \over 2} + \sin {ht \over 2} \over
    \cos {ht \over 2} - \sin {ht \over 2}} \right)^{1/\sqrt{3}}
     d\vec{x}^2 \,.
 }
Here $t$ runs from $-\pi/2h$, where there is a curvature singularity, to $\pi/2h$, which is actually a non-singular infinite future when expressed in terms of proper time.  This solution is not new: to my knowledge it first appeared in \cite{clw}.

The main qualitative difference between the full solution \FullSolution\ and the no-back-reaction approximation \FlatChiPhi\ is that the in vacuum is more difficult to define in \FullSolution, due to the singularity.  Two possibly interesting approaches would be to replace the early time evolution by an appropriate instanton, or to find a CFT formulation in which $\alpha'$ corrections are under control.

The solution \FullSolution\ can be lifted to ten dimensions, and in the cases where it comes from IIA strings or $E_8 \times E_8$ heterotic strings, there is a further lift to eleven dimensions, where the size of the eleventh dimension is controlled by the dilaton.  When the lift is done to Horava-Witten theory, it describes two $E_8$ that start far apart at $t = -\pi/2h$, move swiftly together to a minimum distance at $t=0$, and then move apart again.  This is reminiscent of ekpyrosis, but actually it is quite different: in four-dimensional Einstein frame, there is never any contraction of the scale factor.  There is also no singularity when the branes are at their distance of closest approach.  The reason they don't actually collide is that the four-form field $G_4$ is turned on in such a way as to make the $E_8$ branes repel.  It is notable that in ekpyrosis as envisioned in \cite{kosst}, strings become tensionless at the instant that the branes collide.  It seems extremely likely that string creation is important for understanding how the universe might pass through this singular event.

The point of involving only Neveu-Schwarz fields is that it is more likely that one can perform a worldsheet analysis of such backgrounds.  In fact, the background we have considered resembles a limit of NS5-branes, doubly Wick rotated so that radius is replaced by time.  The main stumbling block to finding a CFT description is that the geometry does not factorize in the way that the NS5-brane geometry does, into a three-sphere part, a linear dilaton background, and flat space.

\section{Conclusions}
\label{CONCLUDE}

Part of the appeal of the discussion in section~\ref{PREHEAT} is that nearly tensionless strings arise in a variety of contexts, on loci in moduli space of real codimension two.  Thus, independent of detailed ideas about how string theory connects to the Standard Model, it is plausible to have non-perturbatively constructed strings become temporarily light enough to play a role in post-inflationary cosmology.  Because we remain ignorant of what string vacuum we might really be living in, it's highly desirable to have some ideas about characteristic features of string theory which may become testable, which are difficult to contrive without invoking strings, and which are common to many backgrounds.  Creation of light strings seems to qualify on all counts.  On the other hand, the experimental constraints on precisely how the universe was reheated seem rather weak, so to really make the idea of post-inflationary string creation testable, one must gain a better idea of how such string creation would affect quantities that will be measured in the near future, like properties of the primordial spectrum of perturbations.

On the formal side, a better understanding of string creation from a worldsheet point of view is needed, and it would also be useful to learn more about the decay of highly excited strings.  One reason that it is particularly important to have some examples of closed string creation described on the string worldsheet is that for bounds like \MuBound\ to be violated, one is usually forced to have string scale curvatures in the background.\footnote{The treatment of nearly tensionless strings in section~\ref{PREHEAT} may be an exception to this.}  So a worldsheet treatment is needed to be sure one has $\alpha'$ corrections under control, even before quantum creation processes are contemplated.  I hope to report on some of these issues in the future.

\section*{Acknowledgments}

I thank the organizers of QTS3 for putting together a stimulating workshop.  This work was supported in part by the Department of
Energy under Grant No.\ DE-FG02-91ER40671, and also by a Sloan Fellowship.

\bibliographystyle{ssg}
\bibliography{QTS3}

\end{document}